\documentclass[fleqn,10pt]{wlscirep} 
\title{ Inverse scattering transform analysis of rogue waves using local periodization procedure}
\usepackage{amsmath} \usepackage[normalem]{ulem}

\author[1,*]{St\'ephane Randoux} \author[1]{Pierre Suret}
\author[2]{Gennady El} \affil[1]{Univ. Lille, CNRS, UMR 8523 - PhLAM -
 Physique des Lasers Atomes et Molécules, F-59000 Lille, France}
\affil[2]{Department of Mathematical Sciences, Loughborough
 University, Loughborough LE11 3TU, United Kingdom}

\affil[*]{stephane.randoux@univ-lille1.fr}

\begin{abstract}
The nonlinear Schr\"odinger equation (NLSE) stands out as the
dispersive nonlinear partial differential equation that plays
a prominent role in the modeling and understanding of the wave phenomena relevant
to many fields of nonlinear physics. The question of random input
problems in the one-dimensional and integrable NLSE enters within the
framework of integrable turbulence, and the specific question of
the formation of rogue waves (RWs) has been recently extensively studied in this
context. The determination of exact analytic solutions of the focusing
1D-NLSE prototyping RW events of statistical relevance is now
considered as the problem of central importance. Here we address this
question from the perspective of the inverse scattering transform
(IST) method that relies on the integrable nature of the wave
equation. We develop a conceptually new approach to the RW
classification in which appropriate, locally coherent structures
are specifically isolated from a globally incoherent wave train to be
subsequently analyzed by implementing a numerical IST procedure
relying on a spatial periodization of the object under
consideration. Using this approach we extend the existing
classifications of the prototypes of RWs from standard
breathers and their collisions to more general nonlinear modes
characterized by their nonlinear spectra. 
\end{abstract}

\begin{document}

\flushbottom \maketitle
\thispagestyle{empty}

\section*{Introduction}

There is currently much research interest in the subject of the formation
of rogue waves (RWs). The traditional notion of RWs is related to
rare events of large amplitude that appear unpredictably on the ocean
surface \cite{Pelinovskybook,Onorato:01}. From the optical fiber
experiment performed by Solli {\it et al} in ref. \cite{Solli:07}, it
has been understood that RWs are ubiquitous phenomena 
observable  not only in oceanography but also in many other physical contexts
\cite{Onorato:13}. Although the unique mechanism of the RW
formation cannot be drawn
\cite{Akhmediev:10,Onorato:13,Dudley:14,Armaroli:15}, it is now
understood that the one-dimensional focusing nonlinear Schr\"odinger
equation (1D-NLSE) provides a universal description of a variety of
nonlinear localization effects that are compatible with RW
events \cite{Akhmediev:09,Dudley:14,Armaroli:15}. The best known
analytical models for RWs are solitons on finite background
(SFBs) which represent exact homoclinic solutions of the 1D-NLSE having
the far-field behavior of a finite-amplitude plane wave and at the
same time exhibiting local peak amplitudes compatible with the
threshold definition of rogue events \cite{Akhmediev:09,Akhmediev:09b,Kedziora:13,Akhmediev:09c,Dudley:14}.
Taking specific and carefully designed initial conditions, many SFBs
have now been observed in well-controlled experiments performed in
several physical systems
\cite{Kibler:12,Kibler:10,Chabchoub:11,Bailung:11,Chabchoub:12a,Chabchoub:12b,Frisquet:13,Kibler:15}. 

Taking random initial conditions in wave problems ruled by the 1D-NLSE
is very pertinent to the study of RWs because the randomness
of the initial condition opens the way for the statistical treatment
inherent in any physically realistic RW description
\cite{Dudley:14,Toenger:15,Walczak:15}. The theoretical analysis of random input
problems in integrable equations such as the 1D-NLSE enters within the
framework of integrable turbulence
\cite{Zakharov:09,Agafontsev:15,Randoux:14,Walczak:15,Derevyanko:08}.
Regarding the focusing 1D-NLSE, it has been recently shown that the
statistics of the field that is measured at long evolution time
strongly depends on the statistics of the random initial condition. A
plane wave perturbed with a random small noise has been found to
produce a field eventually assuming the gaussian statistics
\cite{Agafontsev:15}. On the other hand, heavy-tailed deviations from
gaussianity have been observed for random fields having gaussian
statistics at the initial stage \cite{Walczak:15, Suret:16}. 
The important questions related to the relationship between the initial condition and 
the formation of RWs have been recently investigated in the framework of 
the inverse scattering transform (IST) method\cite{SotoCrespo:16}. It has been also shown
that random initial conditions can excite a range of SFBs well
described by exact analytic solutions of the 1D-NLSE
\cite{Akhmediev:09,Akhmediev:09b,Dudley:14,Walczak:15,Suret:16,Toenger:15} and 
that 1D-NLSE RWs can arise from collisions between some
solitons \cite{Akhmediev:09, Frisquet:13,Toenger:15}. 

The question of the identification and classification of NLSE RWs is a current
issue of importance. So far, this question has been mainly
  investigated by using numerical simulations  
\cite{Dudley:14,Toenger:15,Akhmediev:09,Akhmediev:09b,Walczak:15}.
However the first real-time and direct observation of RWs generated from 
the propagation of partially coherent waves in optical fibers has been 
recently reported in ref. \cite{Suret:16}. In these experiments based on 
the time lens technique, breather-like structures such as Peregrine 
solitons (PSs) have been shown to emerge locally from the random background. 
In the majority of the approaches reported so far, 
RW objects are first specifically isolated from random 
wave trains, and their classification relies on fitting procedures in which the profiles
of interest  are locally compared with
well-known analytic SFB solutions of the focusing 1D-NLSE
\cite{Dudley:14,Toenger:15,Akhmediev:09,Akhmediev:09b,Walczak:15, Suret:16}. In this paper, we
propose a conceptually new approach to the characterization of RWs 
that are observed in wave systems ruled by the focusing
1D-NLSE. We introduce a method relying on the integrable nature of
the 1D-NLSE to compute spectral portraits of localized structures by
using the direct scattering transform, which forms an integral part of
the IST method. Our approach extends
the existing rigid identification of RWs with one of the
prototypical exact breather solutions such as Akhmediev breathers
(ABs), Kuznetsov-Ma (KM) solitons, PSs or the
higher-order SFB solutions describing their collisions.  Instead,
we show that RWs observed in random NLSE solutions represent
more general wave forms which may or may not be very close to one of
the prototypical SFBs.
 
The IST is a well-established method for solving nonlinear integrable
partial differential equations. It has been shown recently that the
IST method can provide a new approach to overcome transmission
limitations in fiber communication channels by encoding information in
the nonlinear IST spectrum \cite{Prilepsky:14,Le:15}. Here, we exploit 
the fact that the IST can be used to determine spectral portraits of
localized structures found in some wave trains of interest. These
spectral portraits (for convenience we shall also call them the {\it
 IST spectra}) provide very accurate signatures of the localized
structures and they can be compared to the spectral signatures of
fundamental solitons and SFBs, which are well-known from the IST
theory. Note that the IST has already been introduced as a tool for
nonlinear Fourier analysis of random wave trains
\cite{Boffetta:92,Osborne:91,Osborne:93,Osborne:94}. This tool has
been successfully implemented in several circumstances to determine
the content of random wave trains in terms of nonlinear oscillating
modes. In particular, the IST analysis has been used to analyze the
soliton content in freak (rogue) wave time series \cite{Slunyaev:06}
and more recently, to evidence the presence of soliton turbulence in
shallow water ocean surface waves \cite{Osborne:15}. Regarding the
specific question of the prediction of RWs, previous
numerical computations of IST spectra have shown that the development
of RWs can be statistically correlated with the proximity to
homoclinic solutions of the 1D-NLSE \cite{Islas:05}. 
Recently the process of RW formation has been studied by computing 
\emph{global} IST spectra characterizing multiple random fluctuations 
found inside a box having a large size \cite{SotoCrespo:16}.
Here, we develop
a new \emph{local} approach in which the objects compatible with prototypes of
RWs are specifically isolated from a wave train to be
subsequently analyzed using a numerical IST procedure that relies on a
spatial periodization of the object under consideration. With this
conceptually new approach, we determine the most essential nonlinear
modes composing the RW under consideration and expand the
existing paradigm that observable RWs are necessarily
described by the standard SFB analytic solutions of the focusing
1D-NLSE.

\section*{Inverse Scattering Transform method to compute spectral portraits}

\subsection*{Spectral portraits of some soliton solutions of the 1D-NLSE}

We consider the focusing 1D-NLSE in the form 
\begin{equation}\label{NLSE}
 i \psi_t + \psi_{xx}+ 2 |\psi|^2 \psi=0 \, ,
\end{equation}
where $\psi(x,t)$ is a complex wave envelope changing in space $x$ and
time $t$. In the IST method, the NLSE is represented as the
compatibility condition of two linear equations \cite{Zakharov:72},
\begin{equation}\label{LP1}
Y_x=
 \begin{pmatrix}
-i \xi & \psi \\ \psi^* & i \xi \\
\end{pmatrix} 
Y,
\end{equation}
\begin{equation}\label{LP2}
Y_t=
 \begin{pmatrix}
-2 i \xi^2+i|\psi|^2 & i \psi_x +2\xi \psi \\ i \psi_x^* - 2\xi \psi^*
& 2 i \xi^2 -i|\psi|^2 \\
\end{pmatrix} 
Y,
\end{equation}
where $\xi$ is a complex spectral parameter and $Y(t,x,\xi)$ is a
vector. The spatial linear operator (\ref{LP1}) and the temporal
linear operator (\ref{LP2}) form the Lax pair of Eq. (\ref{NLSE}).

\begin{figure}[ht]
\centering \includegraphics[width=\linewidth]{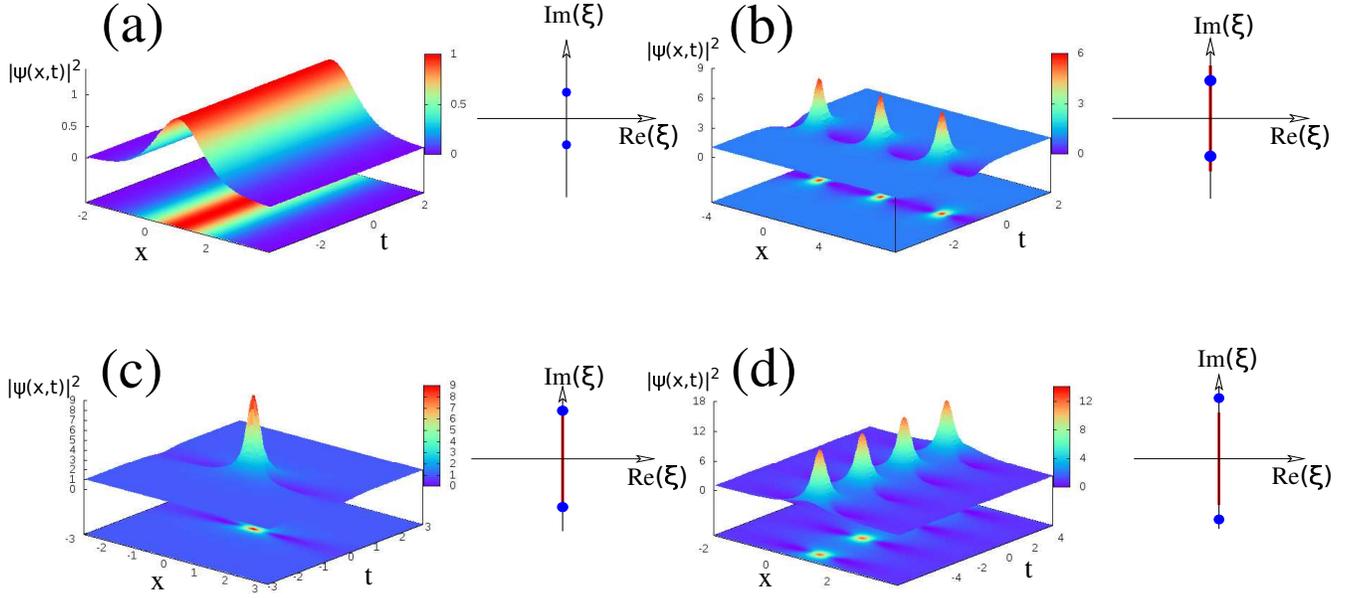}
\caption{{\bf Analytical results from the IST theory}. Spatio-temporal
 evolution (left) and spectral portraits (right) of (a) the
 fundamental soliton, (b) the Akhmediev breather, (c) the Peregrine
 soliton and (d) the Kuznetsov-Ma soliton. The red lines in spectra
 plotted in (b), (c), (d) represent branchcuts. The blue points in
 (a), (b), (c), (d) represent complex conjugate double points.}
\label{fig1}
\end{figure}

For a given potential $\psi(x,t)$ the problem of finding the spectrum
$\{\xi\}$ and the corresponding scattering solution $Y$ specified by
the spatial equation (\ref{LP1}) is called the Zakharov-Shabat (ZS)
scattering problem \cite{yang2010nonlinear}. The discrete eigenvalues
of the ZS operator in (\ref{LP1}) give spectral portraits that provide
precise IST signatures of various solitonic solutions of
Eq. (\ref{NLSE}), which rapidly decay as $|x| \to \infty$. At the same
time, the plane wave solution $\psi =qe^{2iq^2 t}$ has the spectrum
represented by a ``branchcut'' between two points $iq$ and $-iq$ of the
simple spectrum of the {\it periodic} ZS problem
\cite{ma1981periodic,Tracy:84}. This problem can be solved in the
framework of finite-gap theory (FGT) which offers a classification of
periodic and quasi-periodic solutions of Eq. (\ref{NLSE}) according to
their {\it genus}, see Methods. The outlined spectral portraits of
NLSE solitonic solutions determined from the resolution of ZS problem
are shown in Fig. \ref{fig1}. Note that the spectrum of the periodic 
problem also includes the real line, which is not shown in the 
schematic Fig.~\ref{fig1} but will appear in the numerical IST spectrum plots.

As shown in Fig. \ref{fig1}(a), the spectrum of a stable fundamental
soliton $\psi(x,t)=\hbox{sech}(x) e^{it}$ living on a zero-background
is simply made of two doubly-degenerate complex conjugate eigenvalues
$\xi_{\pm}=\pm i/2$. On the other hand, the spectral portraits of
solitons on finite background such as ABs, PSs or KM solitons
essentially represent the spectral portraits of the fundamental
soliton superimposed on the spectrum of the plane wave and differing
only in the relative positions of the soliton and plane wave spectra,
see Fig. \ref{fig1}(b-d) and Methods for the mathematical description
of the spectral portraits of the SFBs. 

\subsection*{Numerical computation of spectral portraits of soliton solutions of the 1D-NLSE}

\begin{figure}[ht]
\centering \includegraphics[width=14cm]{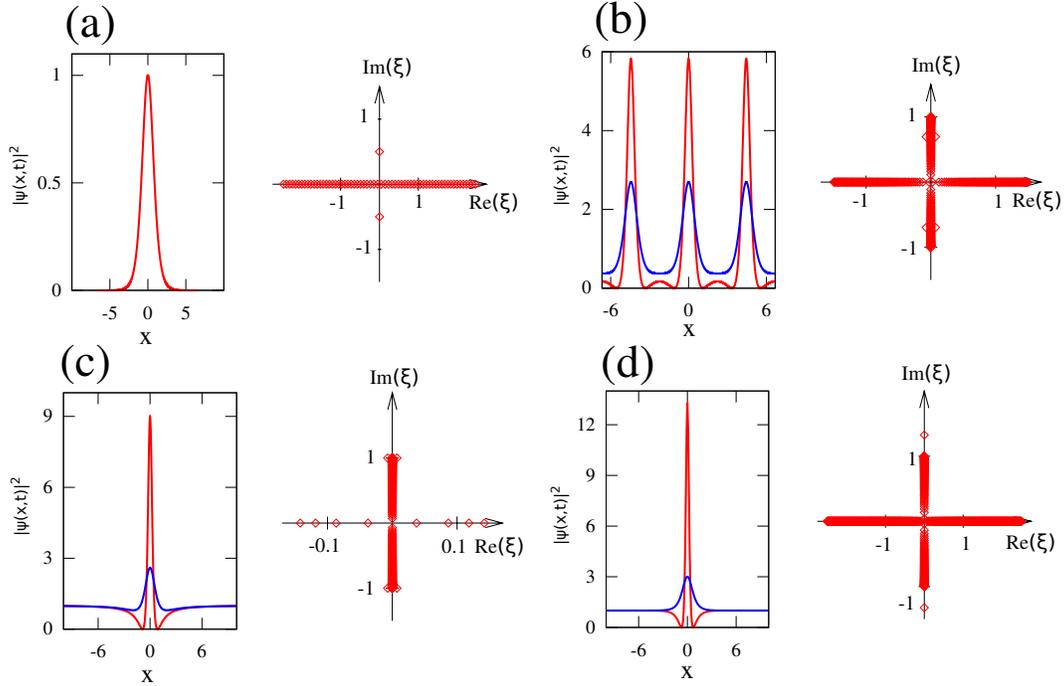}
\caption{{\bf Numerical IST analysis of some soliton solutions of the
  1D-NLSE}. Spatial profiles (left) and spectral portraits (right)
 computed from numerical simulations for (a) the fundamental soliton
 $\psi(x,t)=\hbox{sech}(x) e^{it}$, (b) the Akhmediev breather
 (Eq.(\ref{AB}), $\phi=\pi/4$), (c) the Peregrine soliton
 (Eq.(\ref{AB}), $\phi=0$) and (d) the Kuznetsov-Ma soliton
 (Eq.(\ref{KM}), $\varphi=\pi/4$). The red lines represent power
 profiles $|\psi(x,t)|^2$ at $t=0$ and the blue lines represent power
 profiles $|\psi(x,t)|^2$ at $t=0.5$. Red open squares represent IST
 spectra that are numerically computed both at $t=0$ and at $t=0.5$,
 thus showing that spectral portraits are time-independent. A
 numerical box of size $L=500$ discretized by using $10000$ points
 has been used to compute spectra plotted in (b), (c), (d). }
 \label{fig2}
\end{figure}

Although the spectral portraits of soliton solutions shown in
Fig. \ref{fig1} are given by the IST theory
\cite{Biondini:14,Biondini:15,Gelash:14}, the more general wave
structures are often difficult to analyze, and some numerical
procedures have also been developed to compute IST spectra
\cite{Boffetta:92,yang2010nonlinear}. In our numerical simulations, we
have used a procedure in which Eq. (\ref{LP1}) is rewritten as a
standard linear eigenvalue problem that is subsequently solved by
using the Fourier collocation method (see the
Methods). Fig. \ref{fig2} shows the spectral portraits that are
numerically computed in this way for the fundamental soliton, the AB,
the KM soliton and the PS. 

Some important remarks must be made regarding the use of the numerical
procedure implemented to compute the IST spectra. So far, this
procedure has been proven to be efficient and reliable for the
computation of IST spectra of decaying potentials, such as solitons on
zero background \cite{yang2010nonlinear}. The correct numerical
computation of IST spectra of decaying potentials is achieved when the
size $L$ of the numerical box is significantly greater than the
typical size $\Delta x$ characterizing the decaying potential. The
numerical IST procedure thus provides the complex conjugate
eigenvalues $\xi_{\pm}=\pm i/2$ of fundamental soliton
$\psi(x,t)=\hbox{sech}(x,t) e^{it}$ with a very good accuracy as far
as the size $L$ of the box used for numerical simulations is at least
ten times greater than the typical size $\Delta x \sim 1$ of the
fundamental soliton, see Fig. \ref{fig2}(a). 

To the best of our knowledge, the numerical determination of the
spectral portraits of non-decaying solutions belonging to the family
of ABs, KM and PS has not been made before our work. As for the
fundamental soliton, the IST spectrum of SFB must be computed by
taking a numerical box having a size $L$ greater than the typical
width $\Delta x$ of the SFB. However $L$ must now be {\it much
 greater} than $\Delta x$, to capture the important part of the
spectral portrait related to the non-zero background -- the branchcut, 
or the ``spine'' \cite{tracy1988nonlinear, Osbornebook}, corresponding to 
the spectral band. The spectra plotted in Fig. \ref{fig2}(b-d) have been computed by
taking boxes having a size $L=500$ that is much greater than the
typical width $\Delta x \sim 1$ of the SFB under
consideration. Reducing the size $L$ of the numerical box while
keeping the same number of points used for discretizing the SFB, the
density of spectral points found inside the branch cut region of the
spectrum (i.e. $\xi \in [-i,i] $) decreases. If the size $L$ of the
numerical box becomes comparable to the typical width $\Delta x$ of
the SFB (i.e. $L<10 \, \Delta x$), the branchcut can even be lost and
the spectral signatures numerically determined for the SFB just become
two complex conjugate eigenvalues, as for the fundamental soliton (see also Supplementary Section). 

Note that the IST spectra of SFBs given by Eq. (\ref{AB}) and
Eq. (\ref{KM}) do not depend on time $t$ in agreement with the IST
theory. This is illustrated in Fig. \ref{fig2}(b-d) which shows that
despite the fact that $|\psi(x,t)|^2$ significantly changes between
$t=0$ (red line) and $t=0.5$ (blue line), the IST spectra of the AB,
KM and PS do not change in time.

\subsection*{Spectral portraits of periodized structures}

\begin{figure}[ht]
\centering \includegraphics[width=12cm]{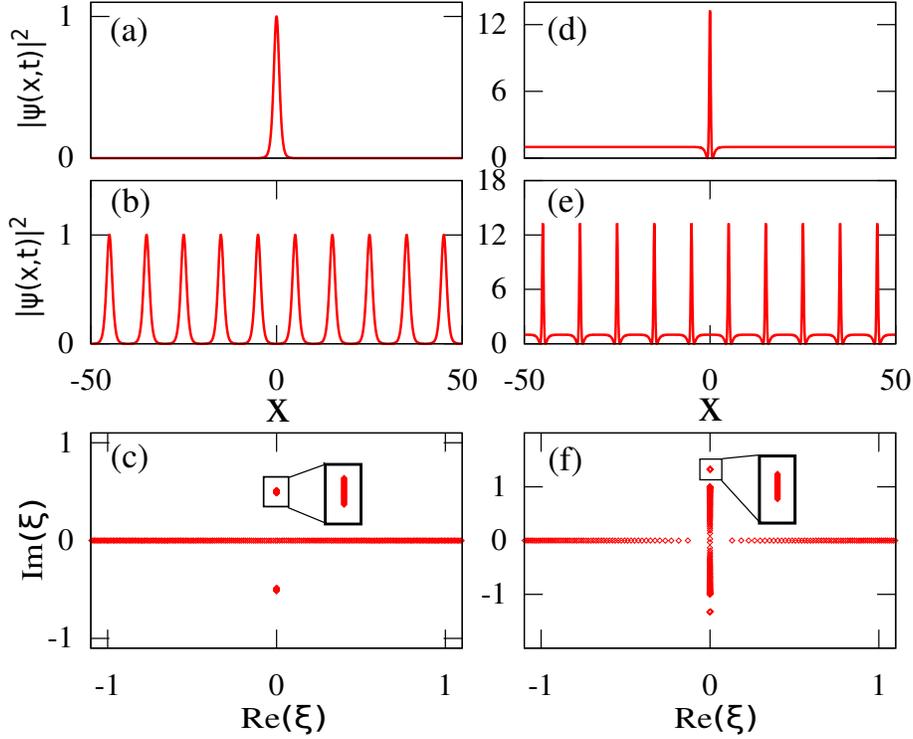}
\caption{{\bf Numerical IST analysis of periodized waveforms} (a)
 Spatial profile of a fundamental soliton $\psi(x,t)=\hbox{sech}(x)
 e^{it}$. (b) Spatial profile of the fundamental soliton periodized
 in space with a period $\Lambda=$10. (c) Spectral portrait of the
 periodized soliton showing that the periodization procedure produces
 a band having a small size. (d) Spatial profile of the KM soliton
 (Eq.(\ref{KM}), $\varphi=\pi/4$). (e) Spatial profile of the KM
 soliton periodized in space with a period $\Lambda=$10. (f) Spectral
 portrait of the periodized KM soliton showing that the periodization
 procedure produces a band having a small size. }
\label{fig3}
\end{figure}

In this paper, we show that numerical IST analysis can be implemented
to get a highly accurate spectral signature of noise-generated
structures that are found in the 1D-NLSE problem with random initial
conditions. However, the implementation of this numerical procedure is
not quite straightforward, and we show in this Section that the
correct determination of IST spectra of localized structures within
more general solutions of the 1D-NLSE requires the IST analysis of
{\it periodic} wavetrains. 

As discussed above, the IST spectrum of SFBs is not qualitatively
properly determined if the size $L$ of the numerical box is comparable
to the spatial width $\Delta x$ of the analyzed SFB (i.e. $L<10 \,
\Delta x$). In other words, the truncation procedure that consists in
performing a {\it local} IST analysis of an isolated SFB amounts to
ignoring the nonlinear interaction between the isolated part of the
SFB and surrounding structures. E.g. by considering one isolated
period of an AB for the numerical IST analysis in a box of the size
$L=2 \pi /p$, we perform the analysis of an isolated object that does
not interact with the neighboring oscillations within the periodic AB
structure. Although this isolated object locally appears to be
practically identical to an AB, the effect of the nonlinear
interaction of the isolated object with the neighbors is lost. As a
result, in these conditions, the numerical IST analysis does not yield
the IST spectrum that is known analytically simply because the
analyzed object is globally not an AB. Thus, the interaction between
the modes is essential for the correct identification of an AB, and
also should be taken into account in the identification of any RW 
object, as also shown in Supplementary Section.

To overcome the fact that a satisfactory numerical IST analysis of
RWs cannot be generally achieved in a local way from a single
isolated object (see also Supplementary Section), 
we introduce here the idea that the spectral portrait
can nevertheless be accurately determined from the IST analysis of an
isolated object, {\it that has been appropriately made periodic in
 space}. The theoretical motivation and numerical justification of this 
idea are presented in the next section  and in the Supplementary Section. 
This is illustrated in Fig. \ref{fig3} that shows that the
IST spectra obtained from the procedure in which localized structures
are truncated to their central core part which is subsequently
repeated to form a periodic function. As shown in
Fig. \ref{fig3}(a)-(c) for the fundamental soliton and in
Fig. \ref{fig3}(d)-(f) for the KM soliton (see Supplementary Section
for the results related to the PS), the IST analysis of
periodic trains that are produced in this way provides the IST spectra
that are very close to the spectra of the pure and non-periodic
objects, see Fig. \ref{fig2}(a) and \ref{fig2}(c). The major
difference between the IST spectra plotted in Fig. \ref{fig2}(a) and
\ref{fig2}(c) and the IST spectra plotted in Fig. \ref{fig3}(c) and
\ref{fig3}(f) lies in the fact that small bands are now found instead
of single points. 

By producing a periodic extension of an isolated localized object, we
realize a local finite-band approximation of the wave field and thus,
no longer ignore the nonlinear interactions between the object and the
surrounding structure. The spatial period $\Lambda$ that is used to
produce the periodic waveform defines the effective intensity of the
interactions, which is translated into the width of the bands in the
IST spectrum whereas the detailed shape of the extracted object
determines the number and location of the bands. The larger the period
$\Lambda$ is, the smaller the bands found in the IST spectrum
are. Therefore the choice of the spatial period $\Lambda$ is crucial
for the quantitatively correct determination of the local IST spectrum
in our numerical procedure. That being said, for spatially isolated
structures such as the genuine fundamental soliton, PS or KM soliton,
the choice of the period of the numerical IST is not essential as long
as the period is much greater than the typical soliton width. Indeed,
our numerically computed spectra for the fundamental soliton and KM
soliton are very similar to the exact IST spectra of the respective
exact solutions shown in Fig. \ref{fig2} as long as $\Lambda$ is
chosen in such a way that the pattern isolated before periodization
includes the soliton part of the SFB together with some part of the
background (see e. g. Fig. \ref{fig3}(f)). The criteria for the
choice of $\Lambda$ providing a robust spectral portrait will be 
discussed in the next Section.

The described procedure has some instructive parallels with the
analysis of dispersive shock waves (DSWs)
\cite{El:2016}: in some cases DSWs can be viewed as
purely solitonic wave trains \cite{conti_observation_2009}, although
generally, they are more accurately represented by the modulated
periodic (genus one) solutions of the relevant equation and exhibit
near-solitonic properties only in the vicinity of one of the
edges. Thus the local IST spectrum a DSW can be captured only by
considering a periodic wavetrain, not a localized pulse, the period
being defined by the distance between the neighboring oscillations.

\section*{Dam break problem and the generation of rogue waves}

\begin{figure}[ht]
\centering \includegraphics[width=12cm]{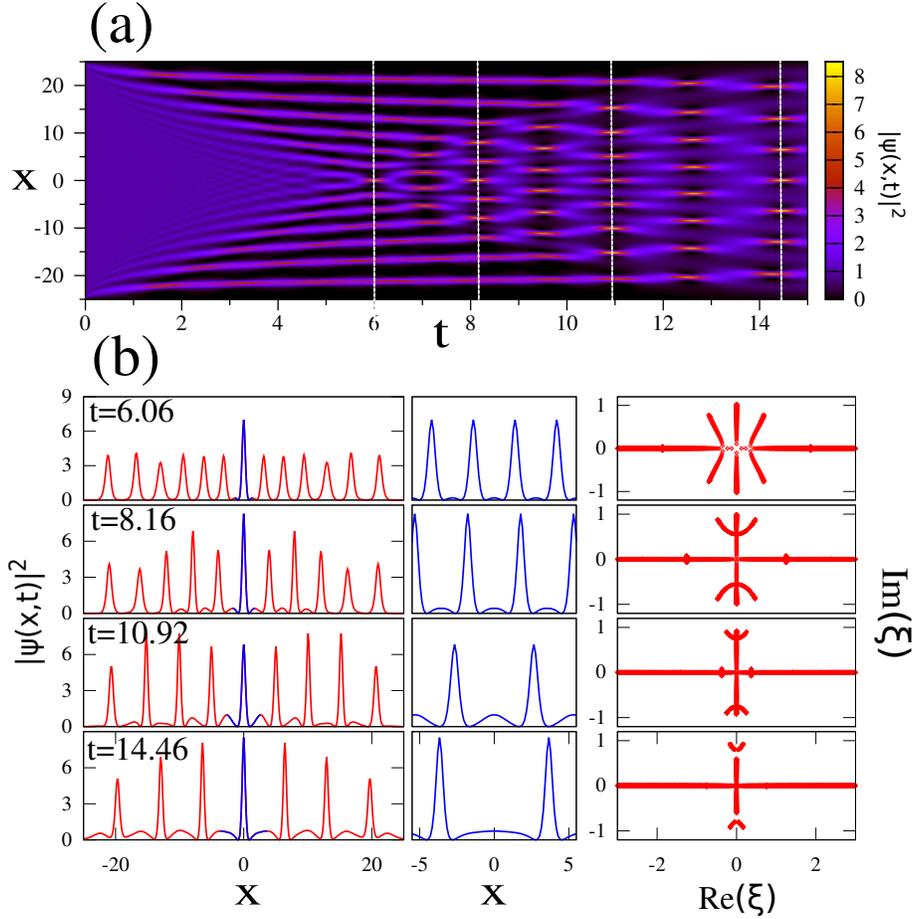}
\caption{{\bf Dam break problem}. (a) Space-time diagram showing the
 evolution of the power $|\psi(x,t)|^2$ of the wave while starting
 from the ``box'' initial condition given by Eq. (\ref{dam_break})
 ($l=25$). Eq. (\ref{NLSE}) is integrated by using a numerical box having 
 a size $L=512$. (b) Numerical IST analysis of periodized waveforms.
 Profiles of the power $|\psi(x,t)|^2$ at times $t=6.06$, $t=8.16$,
 $t=10.92$, $t=14.46$ are plotted in red in the left column. The
 parts of the profiles that are highlighted in blue around $x=0$
 represent the elementary patterns that are periodized to produce
 waveforms shown in the central column. The spectral portraits
 plotted in the right column are computed from the numerical IST
 analysis of the periodic waveforms shown in the central column. The
 numerical IST analysis is made from periodic waveforms including
 $500$ periods. } 
\label{fig4}
\end{figure}

In this Section, we use the tool of numerical IST analysis of
periodized waveforms to investigate the generation of
RWs in the context of the dam break problem recently
considered in ref. \cite{GEl:15}. The dam break problem represents an
analytically tractable scenario of the RW formation in the
framework of the focusing NLSE (\ref{NLSE}). The evolution of an
initial condition having the shape of a rectangular barrier
considered in the small dispersion limit of Eq. (\ref{NLSE}) enables
the generation of the periodic or quasi-periodic nonlinear wave
structures containing many oscillations which can be described within
the semi-classical approximation. During the initial stage of the
evolution these structures are described by the modulated single-phase
(genus one) NLSE solutions and can be associated with DSWs. With the
barrier initial condition, the interaction between two
counter-propagating DSWs has been shown in \cite{GEl:15} to lead to
the emergence of a modulated two-phase large-amplitude breather
lattice whose amplitude profile can be approximated by ABs or PS within certain
space-time regions. More generally, it was shown that the structures 
closely resembling ABs and PSs actually represent modulated two-phase (genus 2) NLS solutions.

The spatio-temporal  diagram plotted in Fig. \ref{fig4}(a) shows the
time evolution of the power $|\psi(x,t)|^2$ while starting from an
initial condition given by: 
\begin{equation}\label{dam_break}
\psi(x,t=0) = \begin{cases} 1 & \quad \text{if } -l<x<l, \\ 0 & \quad
 \text{if} \; \; |x|> l. \\ \end{cases} 
\end{equation}
Note that we solve numerically  a problem with zero boundary
conditions (i.e. the size $L=512$  of the box used for numerical
simulations of Eq. (\ref{NLSE}) is bigger than the size $2l=50$ of the
rectangular barrier).  As shown in Fig. \ref{fig4}(a) and extensively
discussed in ref. \cite{GEl:15}, the DSW collision leads to the
formation of high-power narrow structures localized around
$x=0$. These localized structures observed at $t=6.06$, $t=8.16$,
$t=10.92$, $t=14.46$ are highlighted in blue in the left column of
Fig. \ref{fig4}(b). The spatial size $\Lambda$ of localized structures
that are analyzed by our numerical IST procedure is defined by the
distance separating the maxima reached by the two side lobes
surrounding the localized peak of interest, see blue lines in the left
column of Fig. \ref{fig4}(b). The isolated patterns highlighted in
blue are periodized (see central column in Fig. \ref{fig4}(b)) and the
numerical IST analysis is then made from periodic waveforms including
$500$ periods. Any sufficiently small change in the spatial size
$\Lambda$ will produce IST spectra that are quantitatively slightly
different from those plotted in the right column of
Fig. \ref{fig4}(b). However, as our simulations have shown, the
general result of numerical IST analysis is robust and will not be
qualitatively changed as far as the elementary pattern includes the
peak centered around $x=0$ together with some parts of the side
lobes. The size $\Lambda$ of the elementary pattern characterizes the
effective interaction domain of the central peak with the surrounding
structure. Generally one can propose the following criterion for the
correct choice of $\Lambda$: the period $\Lambda$ is chosen correctly
if any  sufficiently small change in $\Lambda$ produces only small
quantitative change in the spectrum. Some structures (like exact ABs)
could be more sensitive to the variations of $\Lambda $ than others.

The right column of Fig. \ref{fig4}(b) shows the spectral portraits of
``rogue-like'' peaks emerging from the dam break scenario. All the IST
spectra reveal the presence of $3$ main spectral bands, thus
confirming that the observed structures represents genus $2$ solutions
of the 1D-NLSE \cite{GEl:15}. The localized structures observed around
$x=0$ at $t=8.16$ and at $t=14.46$ are very close to the PS in the
sense that they can be locally very well fitted by a profile given by
Eq. (\ref{AB}). However the numerical IST analysis made at $t=8.16$
and $t=14.46$ reveals that those localized structures represent
non-degenerate genus $2$ solutions of Eq. (\ref{NLSE}) that are not
identical to the PS (compare IST spectrum of Fig. \ref{fig2}(c) with
IST spectrum of Fig. \ref{fig4}(b)). All the above features fully
agree with the analytical results of ref. \cite{GEl:15} supporting the
effectiveness of our numerical approach.

It should be stressed that the rectangular barrier
problem can be, in principle,  solved using the classical IST method
with zero boundary conditions. In the  problem with  a wide initial
barrier (or, equivalently, with small dispersion parameter --- see
\cite{GEl:15}) the exact, global, IST  spectrum has  both discrete
and continuous component with a large number of discrete eigenvalues
concentrated along the imaginary axis and  {\it remaining constant
in time}.   However, while this spectrum  implies  a long-time
asymptotic outcome dominated by a large number of fundamental
solitons, it says little about the nonlinear wave field at
intermediate times.  The appearance of the finite-band dynamics,
locally approximating the exact solution's behavior at intermediate
times,  is the result of a complex nonlinear interaction between the
the ``elementary  IST modes''.   The genus of the ``effective''
finite-band potential, appearing as a result of this interaction, as
well as the location and size of the  spectral bands, are the
definitive parameters, which, in particular, characterize proximity
of the observed RW structures to the classical SFBs.

\section*{Noise-driven modulational instability and the generation of rogue waves}

In this Section, we use the tool of the numerical IST analysis of
periodized waveforms to determine the nature of localized structures
that are found in the context of the so-called noise-driven
modulational instability (MI)
\cite{Toenger:15,Agafontsev:15,Dudley:14}. The theoretical
description of the nonlinear stage of MI is now a challenging
question of fundamental
importance\cite{Zakharov:13,Agafontsev:15}. Extensive numerical
simulations have shown that coherent structures localized in space and
time may emerge from noise through the process of MI that is
initiated by a random perturbation of an initial plane wave
\cite{Akhmediev:09,Akhmediev:09b,Toenger:15,Agafontsev:15,Dudley:14}. The question
of the identification and of the characterization of these localized
structures has recently received special attention in the context of
RW generation. Using fitting procedures, it has been shown
that some of these localized structures can be {\it locally}  well
approximated by analytic SFB solutions given by Eq. (\ref{AB}) or by
Eq. (\ref{KM}) \cite{Akhmediev:09,Akhmediev:09b,Dudley:14,Toenger:15,Walczak:15}.

We implement here the numerical IST analysis to get accurate spectral
signatures of some typical noise-generated structures that are found
in the 1D-NLSE problem with random initial conditions. Our study shows
that those localized structures correspond to a variety of {\it
 non-degenerate} genus $2$ and genus $4$ solutions of
Eq. (\ref{NLSE}) that differ from the {\it degenerate} genus $2$
solutions given by Eq. (\ref{AB}) or by Eq. (\ref{KM}). The proposed
numerical IST procedure thus provides a new insight into the
characterization of the RWs found in random wave trains.

\begin{figure}[ht]
\centering \includegraphics[width=12cm]{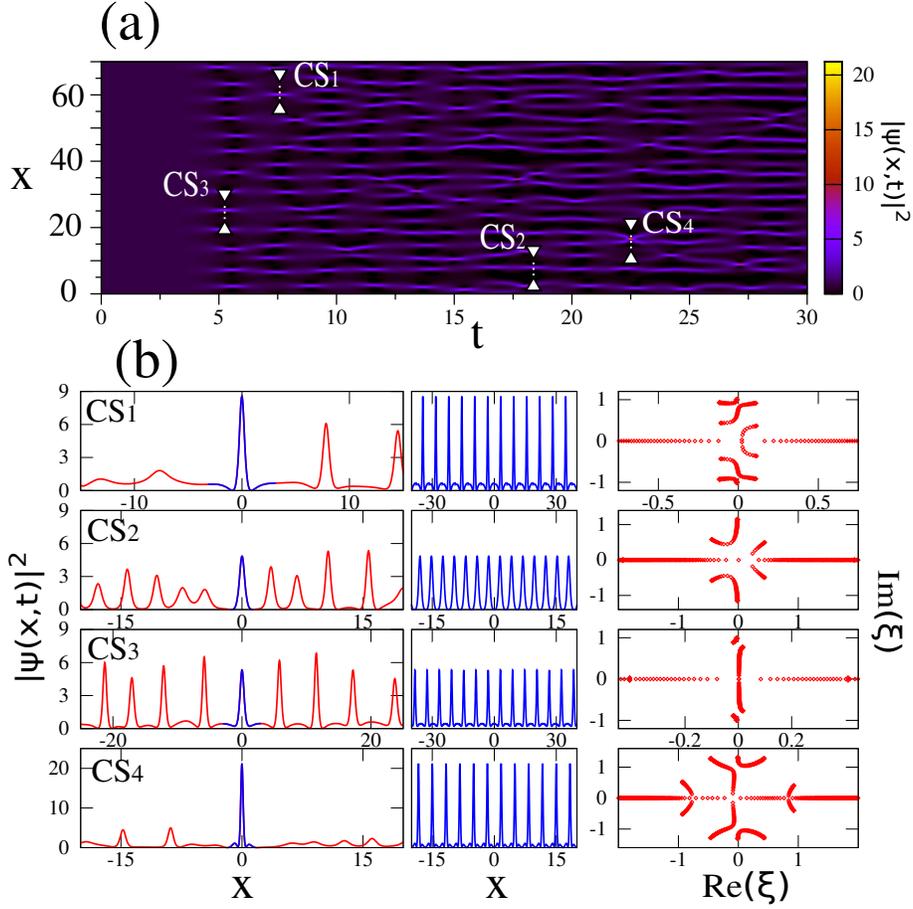}
\caption{{\bf Noise driven modulational instability}. (a) Space-time
 diagram showing the evolution of the power $|\psi(x,t)|^2$ of the
 wave while starting from the initial condition given by
 Eq. (\ref{noise_mi}). (b) Coherent structures are extracted from
 random profiles fluctuating in space $x$ in specific regions labeled
 $CS_1$, $CS_2$, $CS_3$ in (a). The profiles highlighted in blue in
 the left column represent the basic patterns that are periodized to
 produce waveforms shown in the central column. The spectral
 portraits plotted in the right column are computed from numerical
 IST analysis of periodic signals shown in the central column and
 including $200$ periods. The region labeled $CS_4$ in (a) is a
 region in which a strongly localized and intense peak is
 observed. The bottom row of (b) shows the IST spectrum (right
 column) computed from the periodization of this big peak. } 
\label{fig5}
\end{figure}

The spatio-temporal diagram plotted in Fig. \ref{fig5}(a) shows the
time evolution of the power $|\psi(x,t)|^2$ while starting from an
initial condition given by: 
\begin{equation}\label{noise_mi}
\psi(x,t=0) = 1 + \eta(x).
\end{equation}
$\eta(x)$ is a small complex noise field computed from the inverse
Fourier transform of a broadband spectrum under the assumption of a
random phase process, see Methods. As shown in Fig. \ref{fig5}(a), the
spatio-temporal evolution found from our numerical simulations of
Eq. (\ref{NLSE}) is qualitatively very similar to the one evidenced in
ref. \cite{Dudley:14,Toenger:15}. 

Fig. \ref{fig5}(b) shows the results obtained from the IST analysis of
coherent structures which are found in the regions labeled $CS_1$,
$CS_2$, $CS_3$, $CS_4$ in Fig. \ref{fig5}(a). The first row of
Fig. \ref{fig5}(b) shows that the coherent structure $CS_1$ having a
peak power close to $9$ is a genus $4$ solution of the 1D-NLSE. The
analysis of the periodized signal (central column in
Fig. \ref{fig5}(b)) indeed reveals a IST spectrum including $5$ main
bands (right column in Fig. \ref{fig5}(b)). On the other hand,
coherent structures $CS_2$ and $CS_3$ shown in the second and third
rows of Fig. \ref{fig5}(b) have a IST spectrum made with 3 bands. The
coherent structures $CS_2$ and $CS_3$ extracted from wavetrains
nearly periodic in time and in space (see Fig. \ref{fig5}(a)) are
therefore genus 2 solutions of Eq. (\ref{NLSE}). Although the IST
spectrum of $CS_3$ is concentrated around the vertical imaginary axis,
it is however relatively far from the IST spectra of SFBs given by
Eq. (\ref{AB}) or Eq. (\ref{KM}), see Fig. \ref{fig2}(b)-(d) and 
also Supplementary Section for a discussion about results of best fit approximations
of $CS_1-CS_4$. 

The region labeled $CS_4$ in Fig. \ref{fig5}(a) is a region where a
collision occurs between two SFBs in the $(x,t)$ plane. A large peak
with a maximum intensity of $\sim 22$ is formed as a result of this
collision. Such a localized and intense event has already been observed in numerical simulations 
reported in refs. \cite{Akhmediev:09, Toenger:15} where it has been fitted with a rational
breather of order $2$, a degenerate genus $4$ 
solution of Eq. (\ref{NLSE}). The IST spectrum of the second-order rational breather
is made of $5$ spectral bands that
have collapsed to form the IST spectrum consisting of a branch cut and
two complex conjugate points each having quadruple degeneracy.
The fourth row Fig. \ref{fig5}(b) shows that the
coherent structure $CS_4$ resulting from the collision between two SFB is
a genus $4$ solution of Eq. (\ref{NLSE}) because its IST
spectrum is composed of $5$ main bands. However, $CS_4$ is not a degenerate 
genus $4$ solution of Eq. (\ref{NLSE}) and the analysis
of its IST spectrum allows one to clearly distinguish this high-amplitude
coherent structure from the exact second-order breather solution
considered in ref. \cite{Akhmediev:09,Toenger:15} (see also Supplementary Section 
for results about the best fit of $CS_4$.)

\section*{Discussion and conclusion}

From the theoretical point of view, the 1D-NLSE with periodic boundary
conditions and random initial conditions can be solved analytically in
the framework of the FGT because any periodic solution of
Eq. \ref{NLSE} can be approximated by a finite-band potential
expressed in terms of Riemann theta functions over certain algebraic
curve \cite{Agafontsev:15, Osbornebook}. However, this general
mathematical result is very difficult to implement in practice
because the genus of the solution is too large for initial conditions
involving a large number of Fourier modes with random phases. The
numerical realization of the global FGT analysis thus requires very
significant computing resources \cite{Osbornebook}. In contrast, we
perform a local finite-band approximation of the wave field which
includes only the most essential nonlinear interactions. Our approach
thus provides the effective IST spectra giving an accurate signature
of the nature of the isolated pulse and brings a new insight into the
problem of the characterization of RWs and the mechanisms
leading to their formation in integrable turbulence
\cite{Walczak:15,Agafontsev:15}. 

There has been an extensive work on the emergence of the specific SFBs
described by Eqs. (\ref{AB}) and Eq. (\ref{KM}), and their
higher-order collisons \cite{Akhmediev:09, Toenger:15}. However, the
randomness and the interactions among the structures are the factors
that prevent the emergence of {\it exact } SFBs in integrable
turbulence. The IST spectra enable one to quantify the differences
between specific exact solutions of 1D-NLSE and the observed localized
structures.  Indeed, the localized structures found in the
spatio-temporal evolution plotted in Fig. \ref{fig5}(a) correspond
to a variety of {\it non-degenerate} genus $2$ and genus $4$
solutions of Eq. (\ref{NLSE}) that differ from the particular {\it
 degenerate} genus $2$ solutions given by Eq. (\ref{AB}) or by
Eq. (\ref{KM}) or the degenerate genus 4 solutions corresponding to
higher-order rational breathers \cite{Akhmediev:09}. Although we have
not performed an extensive statistical analysis of the content of
random wave trains, it is very unlikely that exact {\it degenerate}
genus $2$ solutions given by Eq. (\ref{AB}) or by Eq. (\ref{KM}) can
be found in the noise-driven evolution of the focusing 1D-NLSE.

The understanding of the statistics of RWs in integrable
turbulence is an open and complex question
\cite{Agafontsev:15,Walczak:15}. In particular, it has been shown that
the stationary probability density function (PDF) of the field is
gaussian if the initial stage consist of a condensate with additional
noise \cite{Agafontsev:15} whereas the stationary PDF is strongly non
gaussian if the initial stage is a partially coherent wave
\cite{Walczak:15}. Islas and Schober \cite{Islas:05} have proposed to
correlate the occurrence of RWs with the proximity to
homoclinic solutions of the 1D-NLSE. The degree of proximity to
homoclinic solutions is determined in ref. \cite{Islas:05} by some
quantitative measurements over the IST spectrum. In a very 
recent paper \cite{SotoCrespo:16}, another approach has been used and 
the ``global'' IST spectra of random initial conditions have been 
computed to study the appearance of RWs during the NLS evolution. We
stress that, in contrast to the above two works, the aim of the
analysis of our paper is not to predict the RW occurrence but
rather to perform an accurate local characterization of coherent
structures already existing in a globally incoherent nonlinear wave
field. As shown in ref. \cite{Bertola:16}, the maximum amplitude 
of a finite-gap solution to the focusing 
1D-NLSE with given spectral bands does not exceed half of the sum of 
the length of all the bands. Using this criterion, the IST spectrum
of a local periodized coherent structure can be used to measure 
the maximum amplitude possibly reached by this coherent structure.
Moreover, our approach could be combined in the future with IST based
predictive methods to obtain a statistical treatment of NLSE rogue
waves.

\section*{Methods}

\subsection*{Mathematical expressions describing the AB, PS and KM soliton and their spectral portraits}

ABs correspond to solutions of
Eq. (\ref{NLSE}) that are periodic in space but localized in time. The
AB solution of Eq. (\ref{NLSE}) can be expressed in terms of a single
real parameter $\phi$ : 
\begin{equation}\label{AB}
\psi_{AB}(x,t)=\frac{\cosh(\Omega t - 2i \, \phi) - \cos(\phi)\cos(p\,
 x)} {\cosh(\Omega t) - \cos(\phi) \cos (p x)} \exp(2it)
\end{equation}
where $\Omega=2 \sin(2\phi)$ and $p=2 \sin (\phi)$.
Fig. \ref{fig1}(b) shows the IST portrait of an AB. The complex
conjugate double points are given by $\xi_{\pm}=\pm i \, \cos(\phi)$
and the endpoints of the branchcut are $\xi_{BC}=\pm i$. The family of
KM solutions corresponds to SFB that are periodic in time $t$ and
localized in space $x$. It can be also expressed in terms of a single
real parameter $\varphi$: 
\begin{equation}\label{KM}
\psi_{KM}(x,t)=\frac{\cos(\Omega t - 2i \, \varphi) -
 \cosh(\varphi)\cosh(q\, x)} {\cos(\Omega t) - \cosh(\varphi) \cosh(q
 x)} \exp(2it),
\end{equation}
where $\Omega=2 \sinh(2\varphi)$ and $q=2 \sinh (\varphi)$.
Fig. \ref{fig1}(d) shows the IST portrait of a KM soliton. The complex
conjugate double points are given by $\xi_{\pm}=\pm i \,
\cosh(\varphi)$ and the endpoints of the branchcut are $\xi_{BC}=\pm
i$, as for the AB. In the limit where $\phi \rightarrow 0$ or $\varphi
\rightarrow 0$, the period of AB and KM solutions tends to infinity
and the solution of Eq. (\ref{NLSE}) that is localized both in space
and time is named Peregrine soliton. In the spectral portrait of the
PS, the complex conjugate double points coincide with the endpoints of
the branchcut ($\xi_{\pm}=\xi_{BC}=\pm i$), as shown in
Fig. \ref{fig1}(c). 

\subsection*{Numerical Simulations}

The determination of discrete eigenvalues $\xi$ of the Zakharov-Shabat
system is made by rewriting Eq. (\ref{LP1}) as a standard linear
eigenvalue problem 
\begin{equation}\label{eigen}
 \begin{pmatrix}
- \partial_x & \psi(x,t) \\ \psi(x,t)^* & \partial_x \\
\end{pmatrix} 
Y=i \xi Y.
\end{equation}
The $x-$axis is truncated into a finite box of size $L$. The
eigenvector $Y=(y_1(x),y_2(x))^T$ as well as the potential $\psi(x,t)$
are expanded into Fourier series with $2n+1$ modes. These Fourier
expansions are substituted in Eq. (\ref{eigen}) and the obtained system for the eigenvalues
is then solved by using standard linear
algebra routines \cite{yang2010nonlinear}. IST spectra plotted in
Fig. \ref{fig2} have been obtained by taking boxes of size $L=500$
that have been discretized by using $10^4$ points. IST spectra plotted
in Fig. \ref{fig3} and in Fig. \ref{fig4}(b) have been obtained from
series including $500$ periods that have been discretized by using
more than $2.10^4$ points. IST spectra plotted in Fig. \ref{fig5}(b)
have been obtained from series including $200$ periods that have been
discretized by using more than $2.10^4$ points.

Numerical simulations of Eq. (\ref{NLSE}) have been performed by using
a pseudo-spectral method working with a step-adaptative algorithm
permitting one to reach a specified level of numerical accuracy. In
Fig. \ref{fig4}, a numerical box of size $L=512$ has been discretized
by using $2^{14}$ points. In Fig. \ref{fig5}, a box of size $L=2000$
has been discretized by using $2^{16}$ points.

The random complex field $\eta(x)$ used as initial condition in
Eq. (\ref{noise_mi}) is made from a discrete sum of Fourier components
: 
\begin{equation}\label{ini_field}
 \eta(x)=\sum\limits_{m} \widehat{X_{m}} e^{i m k_0 x}.
\end{equation}
with $\widehat{X_{m}}=1/L \int_0^L \eta(x) e^{-i m k_0x} dx$ and
$k_0=2 \pi/L$. The Fourier modes
$\widehat{X_{m}}=|\widehat{X_{m}}|e^{i \phi_{m}}$ are complex
variables. We have used the so-called random phase (RP) model in which
only the phases $\phi_{m}$ of the Fourier modes are considered as
being random \cite{Nazarenko}. In this model, the phase of each
Fourier mode is randomly and uniformly distributed between $-\pi$ and
$\pi$. Moreover, the phases of separate Fourier modes are not
correlated so that $\langle e^{i\phi_{n}}e^{i\phi_{m}} \rangle = \delta_{nm}$ where
$\delta_{nm}$ is the Kronecker symbol ($\delta_{nm}=0$ if $n\ne m$ and
$\delta_{nm}=1$ if $n=m$). With the assumptions of the RP model above
described, the statistics of the initial field is homogeneous, which
means that all statistical moments of the complex field $\eta(x)$ do
not depend on $x$ \cite{Picozzi:14}. In the RP model, the power
spectrum $n_0(k)$ of the random field $\eta(x)$ reads as : 
\begin{equation}\label{power_spectrum}
\langle \widehat{X_{n}}\widehat{X_{m}} \rangle =n_{0n} \, \delta_{nm}=n_0(k_n). 
\end{equation}
with $k_n=n\,k_0$. In our simulations, we have taken a random complex
field $\eta(x)$ having a gaussian optical power spectrum that reads 
\begin{equation}\label{gaussian_ci}
n_0(k)=n_0 \, \exp \left[- \left( \frac{k^2}{\Delta k^2} \right)
 \right]
\end{equation}
where $\Delta k$ is the half width at $1/e$ of the power spectrum.
The values of $n_0$ and $\Delta k$ taken in our numerical simulations
are $n_0=5.645. 10^{-3}$ and $\Delta k=0.5$.

\subsection*{Finite-Gap Theory}

The periodic ZS problem is generally solved in the class of the
so-called finite-band potentials which are non-decaying periodic or
quasi-periodic NLSE solutions having the ZS spectrum filling several
bands of finite width \cite{Tracy:84, tracy1988nonlinear,
 Osbornebook}. The finite-band (or as it is often called, finite-gap)
theory (FGT) is widely recognized as a natural framework for the
analysis of nonlinear modulational instability and the formation of
RWs although its practical implementation for the analytic
description of integrable turbulence encounters some fundamental
difficulties \cite{Agafontsev:15}. Within the FGT the multi-phase NLSE
solutions are characterized by a {\it genus}, calculated as $N-1$,
where $N$ is the number of spectral bands. Physically, the genus
characterizes the number of degrees of freedom (i.e. the number of
fundamental oscillatory modes, or phases) within the nonlinear
periodic or quasiperiodic solution for the envelope of the plane
wave. \cite{Osbornebook}. Mathematically, the solution genus
represents the genus of the hyperelliptic Riemann surface, on which
the finite-band NLSE solution is defined in terms of theta-functions
\cite{Osbornebook}. From the viewpoint of the FGT, the plane wave
itself is classified as a regular genus $0$ solution while the
fundamental soliton represents a degenerate genus $1$ solution with
two complex conjugate, doubly-degenerate spectral points of the
periodic problem, the counterparts of the discrete spectrum in the ZS
problem with decaying potentials. The standard SFBs (ABs, KM solitons
and PSs) all are the degenerate genus $2$ solutions. Their spectral
portraits determined from the resolution of ZS problem are shown in
Fig. \ref{fig1}.


\section*{Acknowledgements}

This work has been partially supported by Ministry of Higher Education
and Research, Nord-Pas de Calais Regional Council and European
Regional Development Fund (ERDF) through the Contrat de Projets
Etat-R\'egion (CPER) 2007–2013, as well as by the Agence Nationale de
la Recherche through the LABEX CEMPI project (ANR-11-LABX-0007) and
the OPTIROC project (ANR-12-BS04-0011 OPTIROC). This work has been
also partially supported by the Interuniversity Attraction Poles
program of the Belgian Science Policy Office, under grant IAP
P7-35. The authors thank A. R. Osborne, A. Tovbis, T. Grava and
E. Khamis for fruitful discussions. S. R. thanks J. P. Flament,
F. R\'eal and V. Vallet from Laboratoire PhLAM for technical
assistance with computer ressources.

\section*{Author contributions statement}

All the authors have conceived the original idea. S. R. and P. S.
have performed the numerical simulations. G. E. has interpreted the
results in the framework of finite-gap theory. All the authors have
contributed to the writing and the review of the manuscript. 

\section*{Additional information}

\textbf{Competing financial interests} The authors declare no
competing financial interests.


\end{document}